\documentclass[prb,amsmath,amssymb,showpacs,preprint]{revtex4}
\usepackage{graphicx}
\def\be{\begin{equation}}
\def\ee{\end{equation}}
\def\ep{\epsilon}
\def\ps{\tilde\psi}
\def\N{\mathcal{N}}
\def\var{\langle n^2\rangle}

\begin{document}
\title{Characterization of the spontaneous symmetry breaking due to quenching of a one-dimensional superconducting loop}
\author{Jorge Berger}
\affiliation{Department of Physics, Ort Braude College, 21982 Karmiel, Israel} 
\email{jorge.berger@braude.ac.il}
\begin{abstract}
We study the final distribution of the winding numbers in a 1D superconducting ring that is quenched through its critical temperature in the absence of magnetic flux. The study is conducted using the stochastic time-dependent Ginzburg--Landau model, and the results are compared with the Kibble--Zurek mechanism (KZM). The assumptions of KZM are formulated and checked as three separate postulates. We find a characteristic length and characteristic times for the processes we study. Besides the case of uniform rings, we examined the case of rings with several weak links. For temperatures close or below $T_c$, the coherence length does not characterize the correlation length. In order to regard the winding number as a conserved quantity, it is necessary to allow for a short lapse of time during which unstable configurations decay. We found criteria for the validity of the 1D treatment. The is no lower bound for final temperatures that permit 1D treatment. 
For moderate quenching times $\tau_Q$, the variance of the winding number obeys the scaling $\langle n^2\rangle\propto\tau_Q^{-1/4}$, as predicted by KZM in the case of mean field models; for $\tau_Q\alt 10^5\hbar /k_BT_c$, the dependence is weaker. We also studied the behavior of the system when fluctuations of the gauge field are suppressed, and obtained that the scaling $\langle n^2\rangle\propto\tau_Q^{-1/4}$ is obeyed over a wider range. 
\end{abstract}
\maketitle
\section{Introduction}
The Kibble--Zurek mechanism\cite{K,Z1,Z2} (KZM) aims at the description of kinetically governed phase transitions induced by rapid decrease of the temperature. KZM may be didactically expressed as the division of the transition into three stages, with underlying postulates for each stage. During the initial stage the system is in a high symmetry phase; it explores lower symmetry states due to thermal fluctuations, but is symmetric on the average. The system is able to follow the change of the temperature and may be considered to remain in thermal equilibrium. During the intermediate stage the critical temperature is crossed. During this stage the system is sluggish, so that its final state will be very similar to the initial state. In the final stage several lower symmetry states are long lived; the selected state will have symmetry properties that are determined by the state that was randomly occupied by the system when it entered the second stage.

KZM has been tested by many experiments and simulations, and various degrees of agreement have been obtained. We will be interested in the case of systems with loop topology.\cite{Lag,Carmi,Boris,Tafuri,me,Monaco,Arnab} In this paper we consider a superconducting loop cooled at a fast rate from slightly above to slightly below its critical temperature, in the absence of applied magnetic field, and investigate the possibilities for spontaneous emergence of a permanent supercurrent.

\section{Quantitative Formulation of KZM Postulates}
The first postulate specifies the temperature at which the system becomes sluggish:
\newtheorem{kp}{Postulate}
\begin{kp}
The crossover from the initial to the intermediate stage occurs when $\tau (\epsilon )=|\epsilon /\dot{\epsilon}|$. Here $\epsilon =(T_c-T)/T_c$ is the reduced distance from the critical temperature, $\dot{\epsilon}$ is the derivative of $\epsilon $ with respect to time, and $\tau (\epsilon )$ is the relaxation time of the system.
\end{kp}

The following postulates are formulated here for the case of a system with loop geometry, described by a complex order parameter. The moment at which $\ep =0$ will be taken as $t=0$ and the moment at which the system passes from the initial to the intermediate stage will be denoted as $t=-\hat{t}$.

\begin{kp}\label{P2}
We denote by $\hat{\xi}$ the coherence length of the order parameter at $t=-\hat{t}$ and by $C$ the perimeter of the loop. The behavior of the system can be estimated by enviewing it as if it were divided into $\N=C/\hat{\xi}$ pieces, such that in each piece the phase of the order parameter is uniform, and such that there is no correlation between the phases in different pieces.
\end{kp}

Let us denote by $\theta_i$ the phase difference between regions $i$ and $i+1$. It follows from Postulate~\ref{P2} that the variance of this phase difference is $\langle\theta_i^2\rangle=\pi^2/3$. Let $\theta_C=\sum_{i=1}^\N\theta_i$ be the phase difference accumulated around the loop. Since $\langle\theta_i\theta_j\rangle =0$ for $i\neq j$, if we ignore the constraint that $\theta_C$ has to be an integer multiple of $2\pi$, we obtain
\be
\langle\theta_C^2\rangle=\N\pi^2/3 \;.
\label{vartetc}
\ee
It turns out that the constraint has no effect. Taking the $\theta_i$'s uncorrelated for $i=1,\dots,\N-1$ and picking $\theta_\N$ as the phase difference with the smallest absolute value such that $n=\theta_C/2\pi$ is integer, Eq.~(\ref{vartetc}) is recovered for $\N\ge 3$. $n$ is the winding number of the order parameter, and in view of Eq.~(\ref{vartetc}) obeys
\be
\langle n^2\rangle=\N/12 \;.
\label{varn}
\ee
  
\begin{kp}
The winding number can be regarded as a topological invariant that remains unchanged for $t\ge -\hat{t}$.
\end{kp}
This postulate is far from plausible. Although $n$ is a discrete variable, there is no impediment for a continuous passage of the order parameter to a different winding number, provided that it vanishes at some point. 

The idea behind Postulate 3 is that during the period $-\hat{t}\le t\le \hat{t}$ the system is sluggish and its order parameter practically does not change, and after $t=\hat{t}$ there is an energy barrier between states with different winding number. This statement would be true if at $t=\hat{t}$ the order parameter were close to local equilibrium, which most probably will not be the case. Since the order parameter is random and nearly vanishes at $t=-\hat{t}$, the same is true at $t=\hat{t}$, and we have no good reason to expect a considerable energy barrier for the passage to a different winding number. 

Postulate 3 has been amended in later publications (e.g.\ Ref.~\onlinecite{Arnab}). The revised claim is that (i) there is a time $t_A$ at which the order parameter shoots up sharply; (ii) $n$ remains constant after $t_A$; (iii) $t_A\propto\hat{t}$ and $t_A$ is of the order of $\hat{t}$; (iv) since the coherence lengths are the same for $\ep$ and $-\ep$, it follows from (i)--(iii) that the scaling properties of the system will be the same as if Postulate 3 were true.

Most of the literature on KZM studies the scaling of the density of defects (the winding number in our case) with the rate at which the system is cooled. For this purpose, Postulate 1 is of central importance. In the present study we will first consider the case in which the system is cooled at an ideally fast rate and will check the applicability of Postulates 2 and 3; verification of Postulate 1 will be postponed to Sec.\ \ref{rate}.

\section{Our Model and System}
A simple approach for the description of the dynamics of a superconducting sample is the time-dependent Ginzburg--Landau model (TDGL), with the addition of Langevin terms that bring thermal fluctuations into account (e.g.\ Ref.~\onlinecite{Enamo}); in the case of 1D systems, simplifications are possible.\cite{Lang} We assume that the magnetic field induced by the current around the loop is negligible and write the Ginzburg--Landau energy as
\be
G=\frac{\hbar ^2}{2m}\int_C \left[-\frac{\ep}{\xi^2(0)}|\psi|^2+\beta '|\psi|^4+\left|\left(i\frac{\partial}{\partial s}-\frac{2\pi A}{\Phi_0}\right)\psi\right|^2\right]wds \;,
\label{energy}
\ee
where $m$ is the mass of an electron pair, $s$ is the arc length and the integral covers the loop, $\xi (0)$ is the coherence length at $T=0$, $\psi$ is the order parameter, $A$ is the tangential component of the electromagnetic vector potential, $\Phi_0$ is the quantum of flux, $w$ is the cross section of the loop and $\beta '=4\pi e^2\kappa^2/mc^2$, where $e$ is the electron charge, $\kappa$ is the Ginzburg--Landau parameter and $c$ the speed of light. The usual coherence length in the Ginzburg--Landau model is $\xi (\ep )=\xi (0)/|\ep |^{1/2}$. We assume that $\psi$ and $A$ only depend on time and $s$, and not on the lateral position. $A$ is required in the energy functional in order to take account of the fluctuations of the electric field.

In the numerical procedure, the loop is divided into $N$ segments and the integral is approximated by the sum\cite{Lang}
\begin{eqnarray}
G&=&\frac{\hbar ^2C}{2mN}\sum_{j=1}^N\left(-\frac{\ep_j}{\xi_j^2(0)}|\psi_j|^2+\beta '_j|\psi_j|^4\right)w_j \nonumber \\
 &&+\frac{\hbar ^2N}{4mC}\sum_{j=1}^N\left[\left|\left( \frac{2\pi i C}{N\Phi_0}A_j+1\right)\psi_j-\psi_{j-1}\right|^2
                                          +\left|\left( \frac{2\pi i C}{N\Phi_0}A_j-1\right)\psi_j+\psi_{j+1}\right|^2\right]w_j \;,
\label{discrete}
\end{eqnarray}
where the segment $N+1$ is identified with the segment 1, and the evolution of $\psi_j$ and $A_j$ during a period of time $\Delta t\ll\tau (\ep)$ is given by
\begin{eqnarray}
\psi_j(t+\Delta t)&= &\psi_j(t)-2\Gamma_{\psi,j}\frac{\partial G}{\partial \psi_j^*}\Delta t+\eta_1+i\eta_2 \nonumber \\
A_j(t+\Delta t)&= &A_j(t)-\Gamma_{A,j}\frac{\partial G}{\partial A_j}\Delta t+\eta_A \;.
\label{algo}
\end{eqnarray}
Here $\Gamma_{\psi,j}=NmD_j/\hbar ^2Cw_j$, $\Gamma_{A,j}=Nc^2/\sigma_j Cw_j$, $D$ is the diffusion coefficient and $\sigma $ is the conductivity, $\eta_1$, $\eta_2$ and $\eta_A$ are random numbers with gaussian distribution, zero average and variances 
$\langle \eta_1^2\rangle =\langle \eta_2^2\rangle =2\Gamma_{\psi,j} k_BT\Delta t$, $\langle \eta_A^2\rangle =2\Gamma_{A,j} k_BT\Delta t$, where $k_B$ is the Boltzmann constant.

The material parameters in the model can be evaluated in terms of $T_c$, the Fermi wavevector $k_F$, the mean free path $\ell_e$ and the electron density $n_e$.
Using BCS, dirty limit 
and free electron gas approximations,
\be
\xi^2(0)=\frac{\pi\hbar ^2k_F\ell_e}{12mk_BT_c}\;,\;\;\kappa ^2=0.021\frac{mc^2}{n_e e^2\ell_e^2}\;,\;\;D=\frac{2\hbar k_F\ell_e}{3m}\;,\;\;
\sigma =\frac{n_e e^2\ell_e}{\hbar k_F}\;.
\label{BCS}
\ee

TDGL is not expected to provide a quantitative description of the superconductor dynamics in the entire range to which we will apply it. Nevertheless, it is a self-consistent model from the point of view of statistical mechanics, so that its predictions have at least theoretical value.

An important characteristic length is\cite{Lang} $\xi_\beta=(w\Phi_0^2/32\pi^3\kappa^2k_BT_c)^{1/3}$. For $\ep\approx 0$, $\xi_\beta$ is the order of the length over which $\psi$ is not expected to vanish for a fluctuation such that $G\sim k_BT_c$.

In order to have a 1D situation, $\xi_\beta$ and $\xi (\ep )$ should be larger than the linewith and the thickness. In the absence of significant magnetic fields, the magnetic penetration depth $\kappa\xi (\ep )$ is unimportant.

We note that for the present model the relaxation time does not diverge at $T_c$. First, relaxation of the electromagnetic potential is insensitive to the distance to $T_c$, and indirectly drives the phase of $\psi$. Moreover, for $\ep =0$ the remaining terms in Eq.~(\ref{energy}) lead to a residual relaxation time of the order of $\xi_\beta^2/D$.

\section{Verification of Postulate 2\label{VP2}}
Since this postulate is intended to be applied at a moment such that the system is still able to follow the temperature variation,
in this section we consider systems that are allowed to reach thermal equilibrium at some fixed reduced temperature $\ep =(T_c-T)/T_c$, and check the result predicted by Eq.~(\ref{varn}).

We studied rings of length $C=0.1\,$cm and electron density $n_e=10^{23}$cm$^{-3}$, divided into 300 computational cells. The system evolved during a lapse of time $2\xi_\beta ^2/D$, in steps of $3.3\times 10^{-5}\xi_\beta ^2/D$. At the end of this evolution the winding number $n=(1/2\pi )\sum\arg (\psi_j^*\psi_{j+1})$ was evaluated. For each temperature above (respectively below) $T_c$ this procedure was repeated 800 (respectively 400) times. The evolution time seemed to be sufficient, since there was no significant correlation beween the initial and the final winding number; besides, except for temperatures considerably below $T_c$ and unrealistic initial distributions, longer runs lead to essentially the same results.

\subsection{Uniform loop\label{unif}}
We studied both the case $T>T_c$ and the case $T<T_c$. The initial distribution of the order parameter was taken from a quadratic approximation, as described in the Appendix.
Figure \ref{uniform} shows our results. We observe that (i) $\var$ is a universal function of $\xi$ and $\xi_\beta $; the rhombs and the circles in the figure, that were obtained for samples with different parameters but the same value of $\xi_\beta $, lie along the same curve; (ii) for $T\approx T_c$, $\var$ is surprisingly close to $C/12\xi_\beta $; (iii) for $T>T_c$ and $\xi\ll\xi_\beta $, $\var$ does not depend significantly on $\xi_\beta $; (iv) as the temperature decreases below $T_c$, $\var$ decreases until the probability for $n\neq 0$ becomes negligible.

\begin{figure}
\scalebox{0.85}{\includegraphics{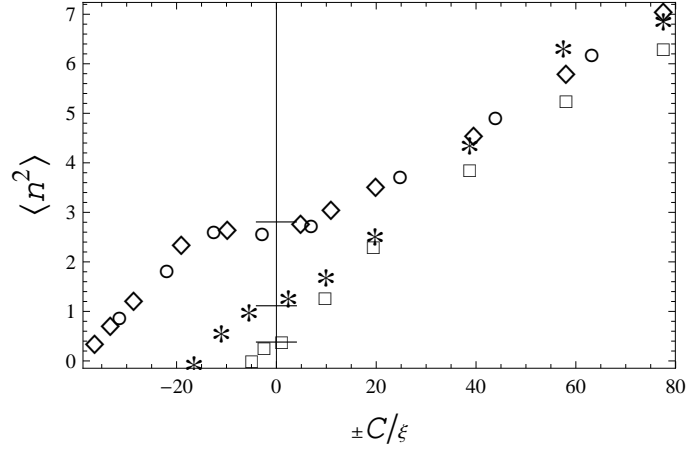}}
\caption{\label{uniform} Variance of the winding number of the order parameter for a uniform ring close to $T_c$. For $T>T_c$ (resp. $T<T_c$) the abscissa is $C/\xi$ (resp. $-C/\xi$), i.e., the abscissa is an increasing function of the temperature. The short horizontal lines that cut the $y$-axis mark the values $C/12\xi_\beta $. $\square$: $w = 5\times 10^{-7}$cm$^2$, $\ell_e = 10^{-5}$cm, $T_c = 1\,$K; $\ast$: $w = 10^{-7}$cm$^2$, $\ell_e = 10^{-5}$cm, $T_c = 5\,$K; $\diamond$: $w = 10^{-7}$cm$^2$, $\ell_e =2.5\times 10^{-6}$cm, $T_c = 5\,$K; $\circ$: $w =2.5\times 10^{-8}$cm$^2$, $\ell_e =10^{-5}$cm, $T_c = 20\,$K.}
\end{figure}

The fact that below $T_c$ $\var$ decreases as $C/\xi (\ep)$ increases indicates that $\xi (\ep)$ is not necessarily the characteristic length over which the phase remains essentially unchanged. The relevant length ought to be the correlation length, which can be defined as follows: we define the autocorrelation function $K(s)=\langle {\rm Re}[\psi^*(s')\psi (s'+s)]\rangle/\langle |\psi |^2\rangle$, and then the correlation length $s_K$ is obtained from a fit $K(s)\approx\exp (-s/s_K)$.

$s_K$ and $\xi(\ep)$ are not equivalent concepts for two reasons. First, $\xi(\ep)$ arises from a competition between the stiffness and the condensation energy, whereas in the case of $s_K$ also the thermal energy and the quartic term enter the competition. More fundamentally, $\xi(\ep)$ is a healing length, which is relevant when $\psi$ has to obey constraints, e.g., it has to vanish at every vortex; in the problem we are considering, the only constraint is periodicity. 

\begin{figure}
\scalebox{0.85}{\includegraphics{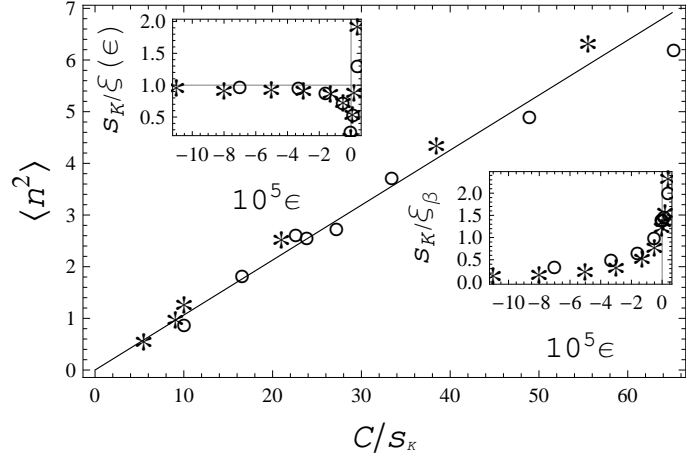}}%
\caption{\label{sk} Variance of the winding number as a function of the ratio between the length of the loop and the correlation length. Insets: correlation length as a function of $\ep =1-T/T_c$. Parameters and symbols as in Fig.~\ref{uniform}.}
\end{figure}

The lower inset in Fig.~\ref{sk} shows the temperature dependence of $s_K(\ep)$, scaled by $\xi_\beta $ (for two samples), and the higher inset compares the temperature dependences of $s_K(\ep)$ and $\xi(\ep)$. Sufficiently above $T_c$, $s_K\approx\xi$. Near $T_c$, $s_K(\ep)$ does not diverge; $s_K(\ep =0)$ is of the order of $\xi_\beta $. Contrary to $\xi(\ep)$, $s_K(\ep)$ is a monotonic function and rises sharply as the temperature decreases below $T_c$. If $s_K$ is not significantly smaller than $C$, then the autocorrelation is influenced by the connectivity of the loop and $K(s)$ is not well fitted by an exponential function.

The main panel in Fig.~\ref{sk} shows $\var$ as a function of $C/s_K$. 
Within our numerical accuracy we obtain a universal line for both samples, which is fitted, both below and above $T_c$, by $\var =0.106C/s_K\approx C/(12\times 0.8s_K)$, i.e., $0.8s_K$ is the effective length over which the phase is essentially uniform.

\subsection{Chain of Junctions}
We now study a loop with $\N_w$ weak links. For a situation in which the phase is fairly uniform between weak links but uncorrelated across them, we could argue that the present case is equivalent to that of Eq.~(\ref{varn}), with $\N =\N_w$. 

We consider a temperature that is above $T_c$ and very close to it.
Between weak links we took $w = 5\times 10^{-7}$cm$^2$, $\ell_e =2\times 10^{-5}$cm, $T_c = 1\,$K, $\ep =-2\times 10^{-7}$; the other parameters were taken as in the previous section. The weak links had lengths $\ell_w$, had a central part with $\ep =-3\times 10^{-2}$ (i.e., had lower $T_c$), and had shorter mean free paths. The outer parts of the weak links had positive values of $\ep$, chosen to compensate the influence that the weak links would have on the rest of the loop due to proximity. This time we took $\psi_j(t=0)=0$. In the evaluation of the winding number, the weak links were skipped.

\begin{figure}
\scalebox{0.85}{\includegraphics{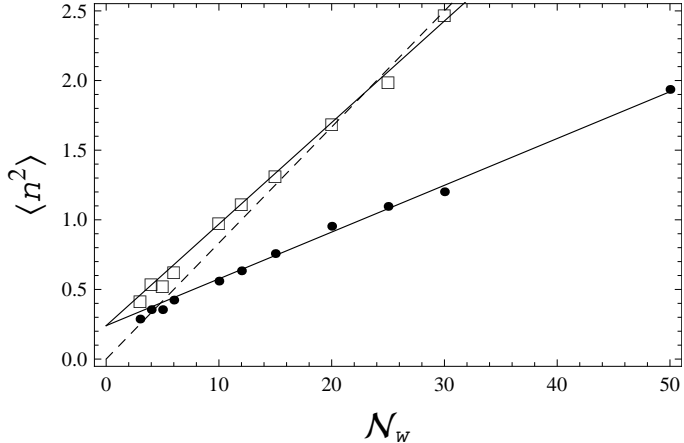}}%
\caption{\label{weak} Variance of the winding number for a loop with $\N_w$ weak links. The dashed line corresponds to $\N =\N_w$. $\bullet$: moderately weak links; $\square$: very weak links. The solid lines corresponds to Eq.~(\ref{varnw}).}
\end{figure}

The dashed line in Fig.~\ref{weak} corresponds to the ideal case $\N =\N_w$. Taking into account that the weak links are not perfect would lead to $\N =p\N_w$, where $p\in (0,1)$ is some ``opacity coefficient." We also have to take into account that, according to the previous section, along the length $C-\N_w\ell_w$ of the strong part of the loop, $(C-\N_w\ell_w)/\xi_\beta $ divisions into different regions effectively occur. Therefore, instead of Eq.~(\ref{varn}) we expect
\be
\langle n^2\rangle=C/12\xi_\beta+(p-\ell_w/\xi_\beta)\N_w/12 \;.
\label{varnw}
\ee

The moderately weak links in Fig.~\ref{weak} had length $\ell_w=1.67\times 10^{-3}$cm, mean free path $10^{-6}$cm, and the region with low $T_c$ had length $3.33\times 10^{-4}$cm; for these weak links we obtained $p=0.45$. The weaker links in the figure had length $3\times 10^{-3}$cm, mean free path $4\times 10^{-7}$cm, and the region with low $T_c$ had length $10^{-3}$cm; for these weak links we obtained $p=0.96$.

\section{Verification of Postulate 3}
The equilibrium distributions that were obtained in Sec.~\ref{VP2} will now be taken as initial distributions, and we will study under what conditions the winding number remains unchanged if the system is {\em instantaneously} cooled to a new temperature $T_f=T_c(1-\ep_f)$ and left at this temperature for a considerable time.

\subsection{Pure 1D case}
Since $\xi (\ep_f)$ may be shorter than the linewidth or the thickness of the loop, our first concern should be whether the system can still be treated as 1D. Following our findings in Sec.~\ref{unif}, we may expect that the relevant requirement is $s_K(\ep_f)\agt w^{1/2}$. We will assume here that the 1D treatment is still applicable; deviations from 1D will be studied in the following section.

It is known that in the absence of magnetic flux a state with winding number $n$ such that $6n^2>C^2/2\pi^2\xi^2(\ep)+1$ is locally unstable (e.g.\ Ref.~\onlinecite{SIAM}), so that Postulate 3 necessarily fails in this situation. In view of Postulate 2 and Eq.~(\ref{varn}), this situation can be avoided provided that $\xi^2(\ep_f)\ll C\hat{\xi}$.

We should require that no further changes of the winding number occur after a state close to a local minimum of $G$ is reached. For this effect, the energy barrier for the passage to another winding number should be significantly larger than $k_BT_f$. It follows from expression (\ref{energy}) that, near a local minimum, the energy per unit length is $-(\hbar^2 w/8m\beta ')(1/\xi^2(\ep_f)-4\pi^2n^2/C^2)^2\approx -\hbar^2 w/8m\beta '\xi^4(\ep_f)$. In order to change the winding number, the order parameter has to vanish at some point and the smallest possible energy increment involved is obtained when the order parameter becomes small in a region of length $\sim\xi(\ep_f)$. From here, $\hbar^2 w/8m\beta '\xi^3(\ep_f)$ has to be significantly larger than $k_BT_f$, a condition equivalent to the requirement that $\xi(\ep_f)$ be significanly less than $\xi_\beta$.

Provided that at time 0 (i) the sample is described by an equilibrium distribution with temperature $T_c(1-\ep )$, (ii) its winding number is $n\neq 0$ and (iii) the temperature is instantaneously lowered to $T_c(1-\ep_f)$, we define the ``conservation probability" ${\mathcal P}$ by
\begin{eqnarray}
{\mathcal P}(\ep ,\ep_f,t)&=&{\rm Probability\;that\;the\;winding\;number\;remains\;invariant}\nonumber\\
&\;&{\rm during\;the\;lapse\;of\;time\;between\;0\;and\;}t\;. \label{P}
\end{eqnarray}
The condition of invariance is required in a strict sense, i.e., $n(t')=n(0)$ for all $0\leq t'\leq t$. According to Postulate 3, ${\mathcal P}$ ought to be 1.

\begin{figure}
\scalebox{0.85}{\includegraphics{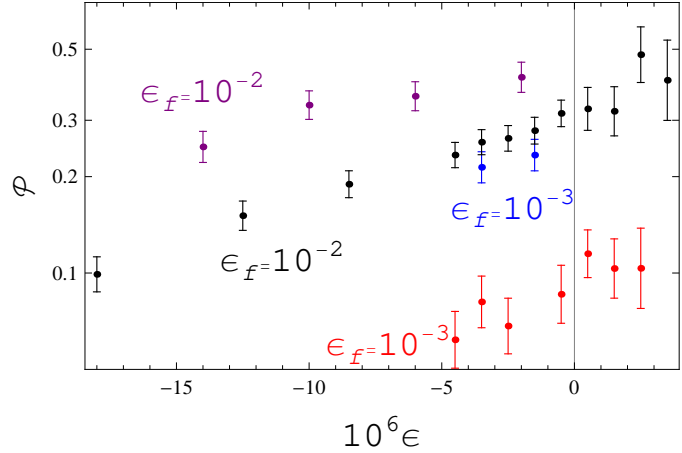}}%
\caption{\label{conserven23}  (Color online) Conservation probability ${\mathcal P}(\ep ,\ep_f,10\xi_\beta^2/D)$ [defined in Eq.~(\ref{P})] as a function of the initial temperature. The red markers are for $\ep_f=10^{-3}$ and the black markers for $\ep_f=10^{-2}$. The blue and the purple markers are obtained if no thermal fluctuations are applied at the final temperature. The error bars were estimated assuming a relative uncertainty equal to 1 over the square root of the number of counts. $w =5\times 10^{-7}$cm$^2$, $\ell_e = 10^{-5}$cm, $T_c = 1\,$K.}
\end{figure}

The markers in Fig.~\ref{conserven23} show the conservation probability for the sample represented by squares in Fig.~\ref{uniform}, as a function of the initial reduced temperature $\ep$. In the case of the black markers, $\xi (\ep_f)\sim 0.02\xi_\beta$, so that the final energy barriers are expected to be considerably larger than $k_BT_f$. We note that ${\mathcal P}$ is substantially less than 1. It also appears that, as the initial temperature increases above $T_c$, ${\mathcal P}$ decreases exponentially down to practically zero within a very narrow range. The red markers are for higher $T_f$ [$\xi (\ep_f)\sim 0.06\xi_\beta$]. It would seem that the energy barriers are still safely high, but the conservation probability for the case $\ep_f =10^{-3}$ is quite lower than in the case $\ep_f =10^{-2}$. In order to estimate the amount by which ${\mathcal P}$ decreases due to thermal jumps over the energy barriers, we repeated some simulations with thermal fluctuations turned off at $t=0$ (blue and purple markers); the absence of fluctuations leads to conservation probabilities that are larger by about $\sim 0.1$. We therefore conclude that thermal fluctuations are not the main reason for the change of the winding number; $n$ usually changes just because the initial order parameter is not close to a local minimum of $G$, and is not within the basin of attraction of the local minimum that has winding number equal to $n(0)$.  
The lower conservation probabilities for small $\ep_f$ could be due to insufficiently fast growth of the order parameter. 

\begin{figure}
\scalebox{0.85}{\includegraphics{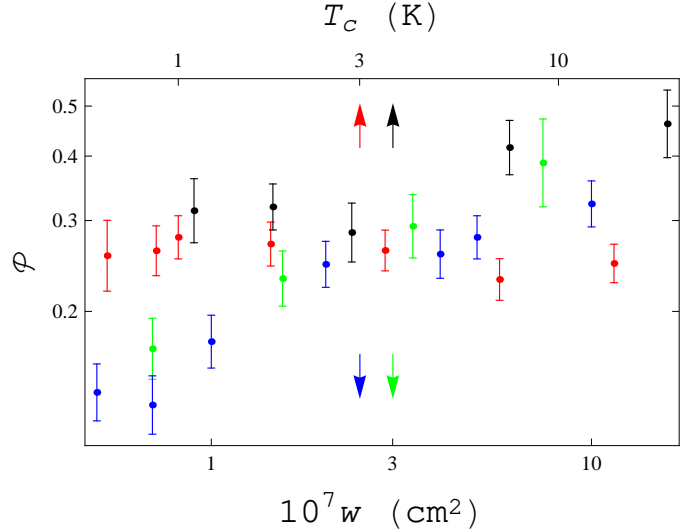}}%
\caption{\label{conswT}  (Color online) Conservation probability as a function of the cross section and of the critical temperature of the wire. 
In all cases $C=0.1$\,cm, $\ell_e = 10^{-5}$cm and $\ep_f=10^{-2}$. When studying the dependence on the cross section (lower $x$-axis), $T_c = 1\,$K; in this study the blue markers are for $\ep =-1.5\times 10^{-6}$ and the green markers for $\ep =1.5\times 10^{-6}$. When studying the dependence on the critical temperature (upper $x$-axis), $w = 5\times 10^{-7}$cm$^2$; in this study the red markers are for $\ep =-1.5\times 10^{-6}$ and the black markers for $\ep =1.5\times 10^{-6}$.}
\end{figure}

Figure \ref{conswT} shows the dependence of ${\mathcal P}$ on $w$ (blue and green markers) and on $T_c$ (red and black markers), for a range somewhat larger than an order of magnitude that includes the sample studied in Fig.~\ref{conserven23}, with initial temperature close to $T_c$. As a general trend, we may regard ${\mathcal P}$ as independent of $T_c$ and proportional to $w^{1/3}$. Extrapolating this result we would obtain ${\mathcal P}(\ep\sim 0 ,10^{-2},10\xi_\beta^2/D)\sim 1$ for $w \sim 10^{-5}$cm$^2$. Close to the parameters in Fig.~\ref{conserven23}, ${\mathcal P}$ is also roughly proportional to $C^{-1/2}$ and to $\ell_e^{1/2}$ (not shown). In addition to the general trends, we have also noticed (but not systematically investigated) a non monotonic detailed structure in the dependence of ${\mathcal P}$ on $\ep$ and on the sample parameters. This dependence may be due to the availability of additional channels when the energy landscape varies.

\subsubsection{A weaker version of Postulate 3\label{weaker}}
At the moment that the sample is (instantaneously) quenched, the order parameter may or may not be within the basin of attraction of the local minimum of the energy with the same winding number. If it is (respectively, is not), we may regard this situation as a sort of ``metastable" (respectively ``unstable") state. We may expect that unstable states decay very fast, so that after waiting a short time the winding number could indeed be conserved. With these heuristic ideas in mind, we define [under the same conditions as Eq.~(\ref{P})] the ``restricted conservation probability" ${\mathcal P}_R$ by
\begin{eqnarray}
{\mathcal P}_R(\ep ,\ep_f,t_u,t)&=&{\rm Probability\;that\;the\;winding\;number\;remains\;invariant}\nonumber\\
&\;&{\rm during\;the\;lapse\;of\;time\;between\;}t_u\;{\rm and}\;t\;. \label{PReq}
\end{eqnarray}
A weak version of Postulate 3 (WP3) is then: {\em for an appropriate short time $t_u$, ${\mathcal P}_R(\ep ,\ep_f,t_u,t)\approx 1$}. 

\begin{figure}
\scalebox{0.85}{\includegraphics{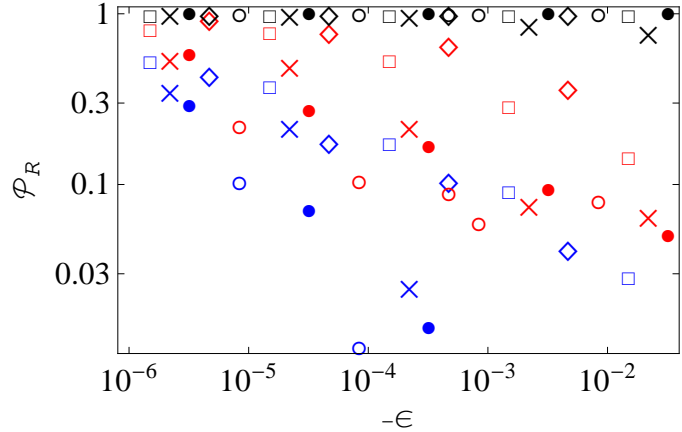}}%
\caption{\label{PR}  (Color online) ${\mathcal P}_R(\ep ,10^{-2},t_u,10\xi_\beta^2/D)$ [defined in Eq.~(\ref{PReq})] as a function of $\ep$, for several samples and several values of $t_u$. The blue markers are for $t_u=10^{-4}\xi_\beta^2/D$, the red markers are for $t_u=10^{-3}\xi_\beta^2/D$ and the black markers for $t_u=10^{-2}\xi_\beta^2/D$. $\square$: $w = 5\times 10^{-7}$cm$^2$, $\ell_e = 10^{-5}$cm, $T_c = 1\,$K, $C=0.1$\,cm;
$\diamond$: $w = 5\times 10^{-7}$cm$^2$, $\ell_e = 10^{-5}$cm, $T_c = 10\,$K, $C=0.1$\,cm;
$\circ$: $w = 5\times 10^{-7}$cm$^2$, $\ell_e = 10^{-5}$cm, $T_c = 1\,$K, $C=1$\,cm;
$\times$: $w = 5\times 10^{-8}$cm$^2$, $\ell_e = 10^{-5}$cm, $T_c = 1\,$K, $C=0.1$\,cm;
$\bullet$: $w = 5\times 10^{-7}$cm$^2$, $\ell_e = 10^{-6}$cm, $T_c = 1\,$K, $C=0.1$\,cm. Uncertainty bars are not shown. }
\end{figure}

Figure \ref{PR} shows the restricted conservation probability as a function of $\ep$ for several samples and choices for $t_u$. The initial temperatures were taken above $T_c$. The black markers are for $t_u=10^{-2}\xi_\beta^2/D$ and the colored markers for shorter waiting times. The squares are for the same sample as the squares in Fig.~\ref{uniform}; all the other markers are for samples in which one of the parameters was changed by a factor 10. Figure \ref{PR} indicates that WP3 is more pervasive than could be expected from a non quantitative statement; except for the sample with small cross section, ${\mathcal P}_R(\ep ,10^{-2},10^{-2}\xi_\beta^2/D,10\xi_\beta^2/D)$ is indistinguishable from 1 in the entire studied range. Note that $t_u$ for the black markers is smaller by three orders of magnitude than the following period of time during which the winding number is monitored for possible changes at the final temperature.

\subsection{Quasi-2D treatment}

\subsubsection{Criterion for one-dimensionality}
Let us model the cross section of the loop as a rectangle of length $\delta $ (i.e., $\delta $ is the width of the loop). How small $\delta $ has to be in order to justify the 1D treatment? It is usually assumed that $\delta $ has to be smaller than the coherence length and the Pearl length. These would be the appropriate requirements if there were a significant magnetic field and the temperature were sufficiently far from $T_c$; we claim that in the present problem the relevant requirement is $\delta\alt s_K$, where $s_K$ is the correlation length defined in Sec.\ \ref{unif}.

It might be argued that $s_K$ was evaluated under the assumption that the system is 1D, so that adopting the same correlation length for the lateral direction is a sort of circular reasoning. As a test for our claim, we consider a simple model in which the lateral direction is taken into account. We divide the loop into two halves, each having length $C$ and width $\delta /2$. We denote the order parameters in the respective halves by $\psi_{(1)}(s)$ and $\psi_{(2)}(s)$ and approximate the lateral component of the gradient of $\psi$ by $2[\psi_{(2)}(s)-\psi_{(1)}(s)]/\delta$. Ignoring the lateral component of the electromagnetic potential, which amounts to ignoring Johnson noise in the lateral direction, the energy becomes
\begin{eqnarray}
G_{\rm Q2D}&=&\frac{\hbar ^2}{4m}\int_C \left\{
\sum_{\nu =1,2}\left[\left(\frac{8}{\delta^2}-\frac{\ep}{\xi^2(0)}\right)|\psi_{(\nu )}|^2+\beta '|\psi_{(\nu )}|^4+\left|\left(i\frac{\partial}{\partial s}-\frac{2\pi A}{\Phi_0}\right)\psi_{(\nu )}\right|^2\right]\right. \nonumber \\
&&\left. -\frac{8}{\delta^2}\left(\psi_{(1)}\psi_{(2)}^*+\psi_{(2)}\psi_{(1)}^*\right)\right\}wds \;.
\label{GQ2D}
\end{eqnarray}

Let us write $\psi_{(\nu )}=|\psi_{(\nu )}|\exp (i\chi_{(\nu )})$ and consider the average $\langle\cos[\chi_{(1)}(s)-\chi_{(2)}(s)]\rangle$ (statistical average, for arbitrary $s$). If $\langle\cos[\chi_{(1)}(s)-\chi_{(2)}(s)]\rangle\approx 1$, it means that the phase variation is practically the same in each of the halves of the loop and the 1D treatment is justified; if $\langle\cos[\chi_{(1)}(s)-\chi_{(2)}(s)]\rangle\approx 0$, there is no correlation between the phase variations in each of the halves, and we have a qualitatively different problem. We note that the energy $G_{\rm Q2D}$ has the same structure as that of a two-band superconductor, and a procedure for the evaluation of $\langle\cos[\chi_{(1)}(s)-\chi_{(2)}(s)]\rangle$ is available.\cite{Milo}

\begin{figure}
\scalebox{0.85}{\includegraphics{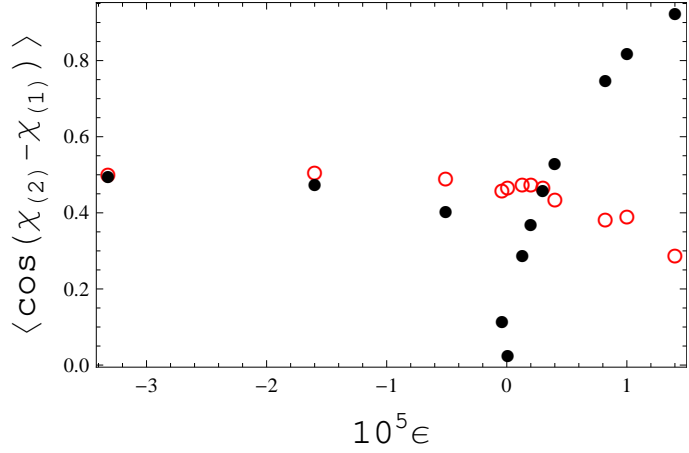}}%
\caption{\label{deltacos}  (Color online) Lateral correlation of the phase of the order parameter. $\circ$: $\delta =s_K$; $\bullet$: $\delta =\xi$. Sample parameters: $C=0.1\,$cm, $w =2.5\times 10^{-8}$cm$^2$, $\ell_e =10^{-5}$cm, $T_c = 20\,$K.}
\end{figure}

Figure \ref{deltacos} shows the temperature dependence of $\langle\cos[\chi_{(1)}(s)-\chi_{(2)}(s)]\rangle$ for the sample represented by circles in Fig.~\ref{uniform}. When we take $\delta =\xi$ (filled symbols), there are temperature ranges where $\langle\cos[\chi_{(1)}(s)-\chi_{(2)}(s)]\rangle\approx 0$, indicating that a loop of width $\xi$ can not be treated as 1D; there are also ranges where $\langle\cos[\chi_{(1)}(s)-\chi_{(2)}(s)]\rangle\approx 1$, indicating that wider samples could still be treated as 1D. On the other hand, taking $\delta =s_K$ (empty red symbols) leads to intermediate correlations for all the considered cases, showing that this is the verge of the 1D behavior. 

There is a moderate tendency for a decrease of $\langle\cos[\chi_{(1)}(s)-\chi_{(2)}(s)]\rangle$ for $\delta =s_K$ as the temperature is lowered, but, in view of our results in the lower inset in Fig.~\ref{sk},
we still conclude that the admissible width for 1D treatment does not decrease as temperature decreases below $T_c$. It follows that if the order parameter is initially 1D, thermalization at $T_f$ will not take the system away from one dimensionality, even if $\xi (T_f)\ll\delta $.

\subsubsection{Influence of the induced flux\label{InflL}}
In the previous sections we entirely neglected the magnetic field induced by the current $I(t)$ that circulates around the loop. For a quasi 1D loop, the influence of this magnetic field can be taken into account by means of a self inductance $L$ that gives rise to a magnetic flux $LcI$ enclosed by the loop. 
Self inductance restricts the rate at which current can vary in a circuit to the characteristic time $\tau_L=L\sigma w/C$. Since usually a change in winding number is accompanied by a change in the current, our results in Fig.~\ref{PR} lead us to naively expect that self inductances such that $\tau_L\agt 10^{-2}\xi_\beta ^2/D$ could result in larger conservation probabilities.

To logarithmic accuracy, $L\sim C/c^2$. From here and Eq.~(\ref{BCS}), $\tau_L\sim 10 (e^2k_B^2T_c^2\ell_e^2n_ew/mc^2\Phi_0^4)^{1/3}\xi_\beta ^2/D$. For the squares in Fig.~\ref{PR} $\tau_L\sim 0.02\xi_\beta ^2/D$, indicating that for quasi 1D samples it would be rather difficult to have $\tau_L$ significantly larger than $\xi_\beta ^2/D$.

We now study the evolution of the flux during a step between $t$ and $t+\Delta t$. We write the flux as $(C/2N)\sum [A_j(t)+A_j(t+\Delta t)]=LcI$ and determine the current by $NI=\sum (I_{Nj}+I_{Sj})$, where $I_{Nj}$ and $I_{Sj}$ are the normal and the superconducting current at segment $j$. The normal current is given by $I_{Nj}=\sigma w[A_j(t)-A_j(t+\Delta t)+\eta_{Aj}]/c\Delta t$ and the superconducting current can be taken as $I_{Sj}=-(2e\hbar wN/mC){\rm Im}(\ps_j^*\ps_{j+1})$, where $\ps (s)=\exp[(2\pi i/\Phi_0)\int_0^s A(s')ds']\psi (s)$ is the gauge invariant order parameter. From here we can obtain
\be
\sum_{j=1}^N A_j(t+\Delta t)=\frac{1}{2\tau_L+\Delta t}\sum_{j=1}^N \left[(2\tau_L-\Delta t)A_j(t)+2\tau_L\eta_{Aj}
-\frac{4e\hbar cN\Delta t\tau_L}{mC\sigma}{\rm Im}(\ps_j^*\ps_{j+1}) \right]\;.
\label{fluxev}
\ee
The flux enclosed by the loop acts a constraint on the values of $A_j$ for each individual segment. It should be noted that the winding number is that of $\psi$, not that of $\ps$.

\begin{figure}
\scalebox{0.85}{\includegraphics{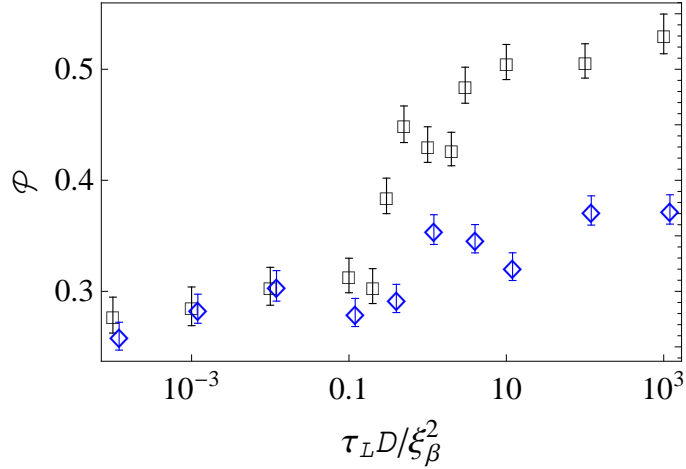}}%
\caption{\label{tauL}  (Color online) Conservation probability ${\mathcal P}(-10^{-6} ,10^{-2},6.7\xi_\beta^2/D)$ as a function of the inductive time constant. The squares (black) and the rhombs (blue) correspond to the same samples as in Fig.~\ref{PR}.}
\end{figure}

Figure \ref{tauL} shows the influence of the induced flux on the conservation probabilities for the samples represented by the squares and by the rhombs in Fig.~\ref{PR}, in the range $0<\tau_L\alt 10^3\xi_\beta ^2/D$. We found that ${\mathcal P}$ does increase with $\tau_L$, but this increase is moderate and ${\mathcal P}$ remains far from 1. There is a limited region of steep increase when $\tau_L$ becomes of the order of $\xi_\beta ^2/D$; the increase in this region is larger for the sample with lower $T_c$.

\section{Influence of the quenching rate\label{rate}}
We finally test Postulate 1. For this purpose we let the sample equilibrate at a temperature considerably above $T_c$, then quench through $T_c$ at a uniform rate $\dot{\ep}=1/\tau_Q$, and finally stabilize at $\ep =\ep_f=0.1$.
Denoting $\hat{\ep}=\ep (-\hat{t})$
, Postulate 1 becomes
\be
\tau (\hat{\ep})/|\hat{\ep}|=\tau_Q \;.
\label{epkova}
\ee
Assuming scalings of the form $\tau =\tau_0/|\ep|^{\nu z}$ and $\xi =\xi_0/|\ep|^\nu$, Eq.~(\ref{epkova}) leads to
\be
|\hat{\ep}|=\left(\frac{\tau_0}{\tau_Q}\right)^{1/(1+\nu z)},\;\;\;\hat{\xi}=\xi_0\left(\frac{\tau_Q}{\tau_0}\right)^{\nu /(1+\nu z)},
\label{scal1}
\ee
so that Eq.~(\ref{varn}) becomes
\be
\langle n^2\rangle\approx (C/12\xi_0)(\tau_0/\tau_Q)^{\nu /(1+\nu z)}\;.
\label{KZn2}
\ee

The initial temperatures were taken in the range $-0.03\alt\ep\alt -0.001$, with $|\ep |\propto\tau_Q^{-1/2}$. The initial values of $|\ep |$ that we took are safely larger than the estimated values of $|\hat{\ep}|$. The region of temperatures covered in this section is much wider than that of the previous sections, and we therefore used variable values for the time step $\Delta t$ and for the number of segments $N$. For $\Delta t$ we typically took $10^{-3}\min [\xi^2(\ep),\xi_\beta^2]/D$. $N$ was initially larger than 300, but when the autocorrelation $K(C/N)$ was steadily greater than 0.9, $N$ was reduced. Equilibration times were typically $2\times 10^3\Delta t$, and changing these times by a factor of 3 had no appreciable influence on our results.

\begin{figure}
\scalebox{0.85}{\includegraphics{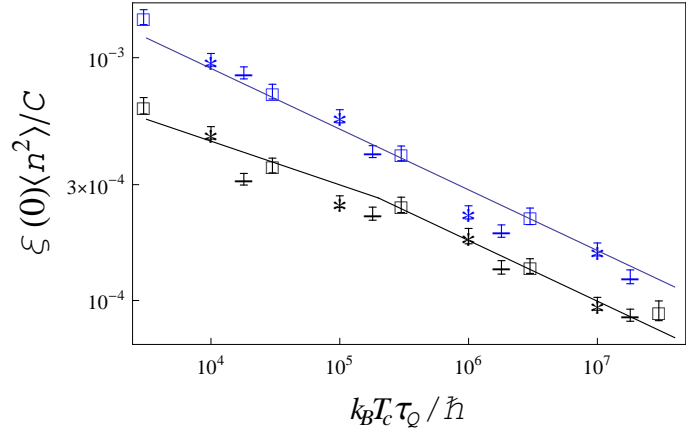}}%
\caption{\label{post1}  (Color online) Variance of the winding number as a function of the quenching time $\tau_Q$. The black symbols were evaluated using the TDGL model; the blue symbols are obtained when electromagnetic fluctuations are ignored. Sample parameters, presented as $(w / 10^{-7}$cm$^2$, $\ell_e / 10^{-5}$cm, $T_c / $K) are $\square$: (5,1,1); $\ast$: (1,0.25,5); $-$: (40,2,32). The statistical uncertainty of $\langle n^2\rangle$ was estimated as $[(\langle n^4\rangle-\langle n^2\rangle^2)/($number of realizations for which the evaluation was repeated)$]^{1/2}$.}
\end{figure}

Our results are shown by black symbols in Fig.~\ref{post1}. 
These results indicate that $\xi (0)\langle n^2\rangle /C$ is a universal function of $T_c\tau_Q$. For mean field, the critical exponents are $z=2$ and $\nu =1/2$, so that the exponent in Eq.~(\ref{KZn2}) is 1/4. We have therefore fitted the rightmost part of the graph to the form $\xi (0)\langle n^2\rangle /C\propto (T_c\tau_Q)^{-1/4}$. Identifying $\xi_0$ with $\xi (0)$, this fit corresponds to $\tau_0\sim 2\times 10^{-5}\hbar /k_BT_c$ in Eq.~(\ref{KZn2}).

At the leftmost part of the graph, $\langle n^2\rangle$ seems to be a weaker function of $\tau_Q$. The fit in this region corresponds to $\xi (0)\langle n^2\rangle /C\propto (T_c\tau_Q)^{-0.18}$. The crossover occurs at $\tau_Q\sim 10^5\hbar /k_BT_c$. 
Weaking of the dependence of $\langle n^2\rangle$ on $\tau_Q$ should actually be expected for sufficiently fast quench since, as we learned in the previous sections, even in the limit of instantaneous quench, there is a considerable ``loss" of metastable states.

We note that in contrast to the previous sections, where our study focused on the region $|\ep |\ll 1$, the relevant time unit in Fig.~\ref{post1} is $\hbar /k_BT_c$ rather than $\xi_\beta^2/D$.

\section{TDGL without the gauge field}
As discussed in Ref.~\onlinecite{Z2}, in a superconducting loop there is a competition between the fluctuations of the order parameter and those of the gauge field, and it is not obvious in general which of them will dominate. If gauge fields are to play a dominant role, the infinite range and the speed with which different sections of the loop can influence one another could prove significant.

We should clarify that, in the case of the 1D loop we are considering, the induced magnetic flux for winding number $n$ is much less than the fluxoid $n\Phi_0$. Moreover, except for the case of Sec.~\ref{InflL}, the magnetic field is completely neglected. The effect of the electromagnetic potential $A$ is not the creation of magnetic fluctuations in space, but rather of fluctuations of the electric field along the loop.

In this section we set $A\equiv 0$ in Eq.~(\ref{energy}), and proceed exactly as before. The variances of the winding number for equilibrium distributions are still described by Fig.~\ref{uniform} and $\langle n^2\rangle\approx C/12\xi_\beta$ close to $T_c$. The probabilities for decays of states with non zero winding number are still fairly described by Fig.~\ref{PR}.

The blue symbols in Fig.~\ref{post1} are our results for $\xi (0)\langle n^2\rangle /C$ as a function of $k_BT_c\tau_Q/\hbar$. As in the case of Sec.~\ref{rate}, we obtain a universal function, independent of the loop parameters. The blue line is a fit to the form $\xi (0)\langle n^2\rangle /C\propto (T_c\tau_Q)^{-1/4}$. As opposed to the case of Sec.~\ref{rate}, the increase of $\xi (0)\langle n^2\rangle /C$ with the quenching rate does not seem to saturate for $\tau_Q\agt 10^3\hbar /k_BT_c$, suggesting that the fluctuations of the electromagnetic field are responsible for this saturation trend.

In the absence of electromagnetic fluctuations the final winding numbers are larger than in their presence. This means that these fluctuations are more effective on destroying metastable states than on creating them.

\section{Conclusions}
We have analyzed the probabilities for final states with permanent currents when a 1D superconducting loop is quenched through its critical temperature in the absence of magnetic flux. The analysis was based on the time-dependent Ginzburg--Landau model with thermal fluctuations.

The predictions of the Kibble--Zurek mechanism (KZM) are obtained in the appropriate limits. Sufficiently above $T_c$, for coherence lengths $\xi (\ep )$ that are short compared to $\xi_\beta$, the phase of the order parameter remains essentially constant over lengths comparable to $\xi (\ep )$. A large fraction of the ``topological defects" that are present immediately after the temperature is abruptly lowered below $T_c$ decay during a lapse of time of the order of $10^{-2}\xi_\beta ^2/D$, but most metastable states that survive this stage conserve their winding number in the following. The requirement of a sort of ``incubation time" before the winding number becomes locked was also noticed in Ref.~\onlinecite{Arnab}. Equation (\ref{KZn2}) is obeyed for $\tau_Q\agt 10^5\hbar /k_BT_c$, with the mean field critical exponents $z=2$ and $\nu =1/2$. For shorter quenching times, the increase of $\langle n^2\rangle$ with the quenching rate is somewhat slower.

When fluctuations of the gauge field are ignored, the values of $\langle n^2\rangle$ increase, and Eq.~(\ref{KZn2}) is obeyed over a wider range of quenching rates. We have not investigated the influence of the self inductance on the dependence of $\langle n^2\rangle$ on the quenching rate.

\begin{acknowledgments}
This research was supported by the Israel Science Foundation, grant No.\ 249/10. Numeric evaluations were performed using computer facilities of the Technion---Israel Institute of Technology. I am grateful to Arnab Das, Francesco Tafuri, Vladimir Zhuravlev and Wojciech Zurek for their answers to my inquiries.
\end{acknowledgments}

\appendix
\section{Quadratic Approximation for the Distribution of the Order Parameter \label{quad}}
We consider a uniform loop and distinguish between the cases above and below the critical temperature.

\subsection{$T>T_c$}
In this case an undemanding approximation can be obtained by neglecting $|\psi |^4$ and $CA_j/N\Phi_0$ in Eq.~(\ref{discrete}). Decomposing into Fourier components, $\psi_j=\sum_{k=1}^N\varphi_k e^{2\pi i jk/N}$, we can write
\be
G=\sum\left({\rm Re}^2 \varphi_k+{\rm Im}^2 \varphi_k\right)E_k \;,
\label{Gm}
\ee
with
\be
E_k=\frac{\hbar ^2w}{m}\left( -\frac{\ep C}{2\xi^2(0)}+\frac{2N^2}{C}\sin^2\frac{k\pi}{N}\right) \;.
\label{Em}
\ee
From here, the equilibrium distributions for ${\rm Re} \varphi_k$ and ${\rm Im} \varphi_k$ are gaussian, with variance $k_BT/2E_k$. 

\subsection{$T<T_c$}
In this case it will be more realistic to assume that the order parameter is close to the ground state, i.e.,
\be
\psi_j=\sqrt{\ep/2\xi^2(0)\beta'}+u_j+iv_j \;,
\label{uiv}
\ee
where $u_j$ and $v_j$ are real and small. Substituting Eq.~(\ref{uiv}) into Eq.~(\ref{discrete}) and keeping terms up to quadratic in $u$ and $v$, we are left with
\be
G={\rm constant}+\frac{\hbar ^2Cw\ep}{Nm\xi^2(0)}\sum u_j^2+\frac{N\hbar ^2w}{2mC}\sum\left[ (u_{j+1}-u_j)^2+(v_{j+1}-v_j)^2\right]\;.
\ee
We introduce now the decomposition $u_j=\sum_{k=1}^N p_k e^{2\pi i jk/N}$, $v_j=\sum_{k=1}^{N-1} q_k e^{2\pi i jk/N}$, with $p_{N-k}=p_k^*$ and $q_{N-k}=q_k^*$; in the decomposition for $v_j$ we do not include a constant term, since it would just lead to multiplication of the order parameter by a uniform phase. Taking $N$ even, the free energy becomes
\begin{eqnarray}
G&=&{\rm constant}+p_N^2E_N^u+p_{N/2}^2E_{N/2}^u+q_{N/2}^2E_{N/2}^v \nonumber\\
&&+\sum_{k=1}^{N/2-1}\left[ ({\rm Re}^2p_k+{\rm Im}^2p_k)E_k^u+({\rm Re}^2q_k+{\rm Im}^2q_k)E_k^v\right]
\end{eqnarray}
with
\begin{eqnarray}
E_k^v=\frac{4N^2\hbar^2w}{mC}\sin^2\frac{\pi k}{N}\;,& E_k^u=E_k^v+\frac{2\hbar ^2Cw\ep}{m\xi^2(0)}& k=1,\dots ,\frac{N}{2}-1 \nonumber\\
E_{N/2}^v=\frac{2N^2\hbar^2w}{mC}\;,& E_{N/2}^u=E_{N/2}^v+\frac{\hbar ^2Cw\ep}{m\xi^2(0)}\;,& E_N^u=\frac{\hbar ^2Cw\ep}{m\xi^2(0)}\;.
\end{eqnarray}
From here, the equilibrium distributions for ${\rm Re} p_k$, ${\rm Im} p_k$, ${\rm Re} q_k$ and ${\rm Im} q_k$ are gaussian, with variances given by the appropriate $k_BT/2E_k^{u,v}$.

\end{document}